\documentclass[12pt]{article}
\usepackage{amsfonts,amssymb,graphicx}
\newtheorem{proposition}{Proposition}
\newtheorem{remark}{Remark}
\newtheorem{lemma}{Lemma}

\def\R{I\!\!R}

\def\N{I\!\!N}

\def\Z{I\!\!\!\!Z}

\def\F{{\cal{F}}}

\def\E{{\rm E}\,}
\def\P{{\rm P}\,}

\def\V{{\rm Var}\,}

\def\t{\theta}

\def\S{\Sigma}
\def\s{\sigma}

\def\vsni{\vskip 0.2cm}
\def\qed{\diamondsuit}

\def\be{\begin{equation}}
\def\ee{\end{equation}}

\def\ds{\displaystyle}

\def\\{\hfill\break}
\def\={{ \; \equiv \; }}

\def\t_N{\tilde{\Z}_N}
\begin{document}
\baselineskip=18pt





\begin{center}
{\LARGE \bf Mean field behaviour of spin systems
with orthogonal interaction matrix\\[27pt]}
{\normalsize \bf Pierluigi Contucci $^{(1)}$, Sandro Graffi $^{(1)}$, Stefano Isola $^{(2)}$ \\[10pt]}
{\normalsize  $^{(1)}$ Dipartimento di Matematica\\
Universit\`a di Bologna,\\
40127, Bologna, Italy\\[15pt]}

{\normalsize  $^{(2)}$ Dipartimento di Matematica e Fisica dell'Universit\`a
di Camerino and \\
Istituto Nazionale di Fisica della Materia,
  62032 Camerino, Italy\\[15pt]}
{\normalsize contucci@dm.unibo.it, graffi@dm.unibo.it, isola@campus.unicam.it}\\[30pt]
\end{center}
\date{}

\begin{abstract}\noindent

For the long-range deterministic spin models with glassy behaviour of
Marinari,
Parisi and Ritort we prove weighted factorization properties of the
correlation
functions which represent the natural generalization of
the factorization rules
valid for the Curie-Weiss case.
\end{abstract}
\\
\vskip 1cm
{\bf Key Words:} deterministic spin glasses, mean field, factorization rules.

\newpage

\section{Introduction and statement of the results}

Mean field models in statistical mechanics are often introduced
to provide a simplification of other more realistic ones. Their
success is based upon the robust physical meaning of the involved
approximation: each part of the system is considered to feel the
action of the remaining ones through a mean effect which decreases
with the total size of the system. The notion of finite cubes immersed
in a $d-$dimensional lattice with Euclidean distance is then replaced by
the complete graph plunged in an infinite dimensional
lattice whose distance
among points decreases uniformly with the size of the graph.

The simplest and
most celebrated among those models is the Curie-Weiss one:  the first
microscopic theory of ferromagnetism was built on its exact solution.
The expression {\it exact solution} has here a peculiarly strong meaning:
not only the free energy density can be computed in closed form in the
thermodynamic limit but also the entire family of correlation functions.
In fact it turns out that once the two point correlation function is
known all the higher order correlations can be computed as powers of the
former after the thermodynamic limit is performed (actually the
computations yields even order correlation functions; the odd ones are all
zero by symmetry in the zero external magnetic field case) . The theory
is said to have  an order parameter, in this case the local
magnetization, and the factorization property of the correlation
functions can be considered the mathematical description of a {\it mean
field behaviour}.

In this paper we study the sine model defined by the Hamiltonian
\begin{equation}
\label{h}
{\cal H}_N(\s)=-\sum_{i < \, j}\,
\frac{1}{\sqrt{2N+1}}\sin{\left(\frac{2\pi ij}{2N+1}\right)}
\s_i\, \s_j \; ,
\end{equation}
or more generally a spin system with orthogonal interaction matrix.

This class of models has been introduced by Marinari, Parisi and Ritort
in \cite{MPR} and subsequently studied in
\cite{DEGGI}. In the sequel we shall refer to them as MPR models.

They probably provide the first example of long range spin models with
non-random interactions with a genuine mean field spin-glass low-temperature phase.
On the other hand the Hamiltonian (\ref{h}) shares with
the Sherrington Kirkpatrick one a mean field
property since the interaction felt by each spin due to the remaining ones
is in the average the same. In SK the local field is a Gaussian variable
with zero average and unit variance. In this case the local field
\be
h_i \, = \, \sum_{j}\,
\frac{1}{\sqrt{2N+1}}\sin{\left(\frac{2\pi ij}{2N+1}\right)}\, \s_j \;
\ee
can be considered in a natural way as a random variable
uniformly distributed over the lattice
${\bf Z}_N:={\bf Z}\,{\rm mod}\, N$ with
zero average and unit variance.

Our main objective is to establish to what extent the mean field property of
the Hamiltonian is reflected on the factorization properties
of the correlation functions.

The main result of the paper is the natural generalization of the factorization rule
valid for the Curie-Weiss case and can be expressed by the following
\begin{proposition}
\label{intro}
For every positive inverse temperature
$\beta$ except, at most, a set of zero Lebesgue measure, and in particular
for every $\beta <1$ the correlation functions
fullfill the following relation:
$$
\biggl|\, \frac{1}{N^2}\sum J_{i,j}J_{l,m}
<\sigma_i\sigma_j\sigma_l\sigma_m>
 - \;\frac{1}{N^2}\sum J_{i,j}J_{l,m}
<\sigma_i\sigma_j><\sigma_l\sigma_m>\, \biggr|\; = \; {\cal O}\left(\frac{1}{N}\right)
$$
where the sums run over non-coincident indices within each expectation.
\end{proposition}
\\
{\bf Remarks}:
\begin{enumerate}
\item  A little algebra after application of translation invariance shows that
the previous formula, when the interaction coefficients are  $\frac{1}{N}$
(Curie-Weiss) yields the well known factorization rule
$<\s_1 \s_2 \s_3 \s_4> \, = \, <\s_1 \s_2>^2$ in the thermodynamic limit.
\item
Higher order relations  can be found involving pairwise $n$-points connected
correlations, with $n\geq 2$. We postpone  a more precise
statement of these results to Sections 2 and 3.
\item
It is perhaps worth mentioning that such an even-type factorization property
is structurally different from the factorization property of pure states (see
e.g. \cite{MPV}, III.1): the first one describes the reduction of the Gibbs
state to the two point correlation function like in the Gaussian case,
while the latter doesn't hold for all the Gibbs states but only for the
extremal ones in which the former can be decomposed. In each of them there is
a complete factorization of the correlation functions when the thermodynamic
limit is performed.
\end{enumerate}
Our strategy is the following: from the study of the fluctuations
of the intensive quantities (basically the energy per particle) we deduce
factorization properties for the correlation functions for our non
translationally-invariant interactions.
Similar results were obtained in \cite{G} and \cite{AC} for SK.
We want to stress the fact that our approach doesn't rely on the
computation of the solution of the model (still not available at least
on rigorous grounds) but only on those bounds over the fluctuation
of the energy coming from equivalence of ensembles (microcanonical
and canonical) ideas. The main technical tool we use is the
property of orthogonality of the interaction matrix: it allows us
to show first the extensivity of the energy and second to produce the
expected $1/N$ bound on the fluctuations.

The paper is organized as follows: in the coming Section 2 we review the
factorization properties of the Curie-Weiss model through an
analysis of the energy fluctuations and of the high temperature expansion.
We emphasize the fact that all the results we present are obtained, including
the existence of the thermodynamical limit, without making use of the exact
solution of the model. In Section 3 we apply the same methods to the
MPR models and we obtain a properly weighted factorization formula.

\section{Remarks on the Curie-Weiss model}
We begin this paper with a full description of the
high-temperature ($\beta <1$) regime of the
Curie-Weiss model of statistical mechanics along with a discussion of
the factorization properties of the correlation
functions which can be obtained in this regime.

The basic setup is a probability space $(\S_N, \F_N,\P_N)$ defined as follows:
the sample space $\S_N$ is the configuration space, i.e. $\S_N=\{-1,1\}^N$ whose elements are
sequences $\s=\s_1\cdots \s_N$ such that $\s_i\in \{-1,1\}$, $\F_N$ is the finite
algebra with $2^{2^N}$ elements and the {\sl a priori} (or {\sl infinite-temperature}) probability measure $\P_N$
is given by
\be
\P_N(C)={1\over 2^N} \sum_{\s \in C} 1.
\ee
We shall consider systems specified by a global pair interaction Hamiltonian
\be
\label{interaction}
{\cal H}_N(\s)=-\sum_{1\leq i<j\leq N}J_{ij}\,\s_i\, \s_j
\ee
where $J=(J_{ij})$ is a
symmetric nonnegative definite $N\times N$ matrix given from the outset.

The simplest example is the so called  Curie-Weiss model, defined by $J_{ij}\equiv 1/N$.
The partition function $Z_N$ at inverse temperature $\beta$ is defined as
\be
\label{partitionfct}
Z_N(\beta) = \sum_{\s \in \S_N} \exp{(-\beta \,{\cal H}_N(\s))}=2^N\,\E_N (e^{-\beta {\cal H}_N}).
\ee
The Hamiltonian for the Curie-Weiss model is then given by
\be
{\cal H}_N(\s) = -{1\over 2N}
\left(\sum_{i}\s_i\right)^2 +{1\over 2}=-{1\over 2N} {\cal M}_N^2(\s)+{1\over 2}
\ee
where
\be
{\cal M}_N(\s) = \sum_{i=1}^N \s_i
\ee
is the total magnetization and the $1/2$ comes from the fact that we are
not allowing self-interactions (there are no terms with $i=j$ in
(\ref{interaction})). In particular
we have the bounds
\be\label{bounds}
 -{1\over 2} (N-1) \leq {\cal H}_N (\s) \leq  \frac{1}{2},
\ee
The ground state $\s^0$ is the state which maximizes the magnetization, i.e.
${\cal M}_N(\s^0)=N$.

An important yet not well known fact about the Curie-Weiss model is that the
existence of its thermodynamical limit can be proved, as a consequence of the
subadditivity of the free energy density, independently of its exact solution
and using only the bounds on the energy.
\begin{proposition}
\label{subadd}
For every positive integer $k$
\be
\frac{1}{kN} \log Z_{kN}\, \le \, \frac{1}{N} \log Z_N   \; .
\ee
\end{proposition}
{\sl Proof.} We show the formula for $k=2$; the general case runs identically.
The main ingredient is a lemma that can be proved by
an easy combinatorial counting argument:
\begin{lemma}
\label{symmetry}
Let $P_N$ be the number of ways in which the set of $2N$ indices
$1, 2, \cdots, 2N$ can be split into two sets of $N$ indices
and denote ${\cal P}_N$ the set of bipartitions, $P_N=|{\cal P}_N|$.
Then the following identity holds:
\be
{\cal H}_{2N} \, = \, \frac{1}{P_{N}}\sum_{p\in {\cal P}_{N}} \left({\cal H}^{l}_{N}(p) \, + \, {\cal H}^{r}_{N}(p)\right) \,
\ee
where $l$ and $r$ stand for the left and right side of the bipartition $p$.
\end{lemma}
Introducing the uniform probability measure ${\cal E}$ on ${\cal P}_N$ and using
the Jensen inequality we may apply the Griffiths symmetrization argument to get
\begin{eqnarray}
Z_{2N} \, &=& \, \sum \exp^{-\beta {\cal H}_{2N}} \nonumber \\
\, &=& \,   \sum \exp^{-\beta {\cal E}({\cal H}^l_N+{\cal H}^r_N)} \\
\, &\le& \, \sum {\cal E}\left[\exp^{-\beta ({\cal H}^l_N+{\cal H}^r_N)}\right] \nonumber \\
\, &=& \, {\cal E}(Z_N^2) \, = \, Z_N^2 \nonumber\qquad \qquad \qquad \qquad\qed
\end{eqnarray}
We  show now how  the theory of equivalence of ensembles can be used to prove
a factorization property of the $4$-points correlation function. From the
equivalence of the canonical and the microcanonical ensemble we know that
the energy density of the Curie-Weiss model $h_N={\cal H}_N/N$
has vanishing fluctuations
in the Gibbs state when the thermodynamic limit is performed. In particular
the quadratic fluctuations for almost all  $\beta$ (but for a set of zero
Lebesgue measure) are ruled by:
\be
<h_N^2>-<h_N>^2={\cal O} \left(\frac{1}{N}\right) \; .
\ee
An easy computation using the explicit form of the Hamiltonian shows that
the former relation becomes, once the limit $N\to \infty$ is considered and the translation invariance
has been taken into account (see also below):
\be\label
{4pts}
<\s_1 \s_2 \s_3 \s_4> \, = \,  <\s_1 \s_2>^2   \;
\ee
which expresses the vanishing of the $4$-points truncated correlations.

This result  comes from general facts about the Gibbs state and  holds
for almost all temperatures. On the other hand it can be improved to hold
for all high temperatures, where one can also get similar results for more general $2n$-points truncated correlations,
with $n\geq 2$.

To this purpose we can compute
the partition function on that regime (see e.g.\cite{Th}) by first decoupling the spins in the
Hamiltonian through the elementary identity:
\be
e^{\,\, a^2\, b}={1\over \sqrt{2\pi}} \int_{-\infty}^\infty \exp \left(-{x^2\over 2}+
\sqrt{2b}\,\, ax\right) dx,
\ee
with the identifications $a={\cal M}_N(\s)$ and $b=\beta/2N$.
Since
\be
\label{coshona}
\sum_{\s \in \S_N} \exp{ \left({\sqrt{\beta\over N}} {\cal M}_N(\s)\,x\right)}=2^N\left[\cosh
\left({\sqrt{\beta\over N}}x\right)\right]^N,
\ee
we obtain
\begin{eqnarray}
\label{logcosh}
Z_N(\beta) &=& 2^N\,{e^{-{\beta\over 2}}\over \sqrt{2\pi}}
\int_{-\infty}^\infty e^{-x^2/2}\left[\cosh
\left({\sqrt{\beta \over N}}x\right)\right]^N\, dx\nonumber \\
&=&2^N\,e^{-{\beta\over 2}}\,\sqrt{N\over 2\pi \beta}
\int_{-\infty}^\infty \exp \left\{N\left[-{y^2\over 2\beta}+ \log \cosh
{y}\right]\right\}\, dy.
\end{eqnarray}
This formula immediately leads to the following result.
\begin{proposition}
\label{cwpf}
For $0\leq \beta <1$ we have
\be
-\beta\; {F(\beta)}\equiv\log Z_N (\beta) =N\log 2\, + G_N(\beta),
\ee
where
\be
G_N(\beta) \nearrow \sum_{k\geq 2} {\beta^k \over 2k} = -{\beta \over 2}- \log
\sqrt{1-\beta}\quad\hbox{as}\quad N\to \infty.
\ee
\end{proposition}
\begin{remark}\label{cw}
Note that if one includes self-interactions in (\ref{interaction}), namely for an Hamiltonian
${\cal H}_N(\s)= - 2N^{-1} \left(\sum_{i}\s_i\right)^2$, the resulting limiting function is
just $- \log \sqrt{1-\beta}$.
\end{remark}
{\sl Proof.}
Using (\ref{coshona}) with
$\displaystyle (\cosh x)^{N}=\sum_{k=0}^{\infty} x^{2k}\sum_{k_1+\cdots
+k_N=k}\prod_{l=1}^{N}\frac{1}{(2k_l)!}$ and observing that the combinatorial
factor is $\displaystyle \left({N\atop k}\right)/(2^kN^k)=1/(2^k k!)+
{\cal O}\left(\frac{1}{N}\right)$ we get (with two successive change of variables $u^2=2x$ and
$y=(1-\beta)x$, and $\beta<1$)  \begin{eqnarray}\label{logcosh3}
Z_N(\beta)
&=&2^N\,e^{-{\beta\over 2}}\,\sqrt{2\over 2\pi }
\int_{0}^\infty
e^{-(x+\frac{1}{2})}\sum_{0}^{\infty}\frac{\beta^k}{k!}x^k x^{-\frac{1}{2}}dx
\, + \,{\cal O}\left(\frac{1}{N}\right) \\ &=&
2^N\,\frac{e^{-{\beta\over 2}}}{\sqrt{1-\beta}}\frac{1}{\sqrt{\pi}}
\, \int_{0}^{+\infty} e^{-y}y^{-\frac{1}{2}}dy
\, + \,{\cal O}\left(\frac{1}{N}\right)
\end{eqnarray}
which gives the theorem with the observation $\Gamma(0)=\sqrt{\pi}$. $\qed$

We will now deal with the limiting behaviour of the energy density and connected correlations at high
temperature deriving some easy consequences of the above result.
For notational simplicity' sake let $<\cdot >$ denote the thermal average
corresponding to fixed
$\beta$ and $N$\footnote{A more consistent notation would be
$\E_{N,\beta}(\cdot )$, so that $\E_N(\cdot ) \equiv \E_{N,0}(\cdot )$.}, i.e. given $A:\S_N\to \R$,
\be
\label{thermal}
<A> \; =\;  {\sum_{\s \in \S_N} A(\s) \, \exp{(-\beta {\cal H}_N(\s))} \over Z_N(\beta) }\cdot
\ee
Define moreover the energy density
\be\label{ed}
h_N(\s) = {{\cal H}_N(\s)\over N},
\ee
which, by (\ref{bounds}), takes values in the interval $\ds [-{1\over
2},{1\over 2N}]$, and consider its distribution function at (inverse)
temperature
$\beta$, \be \label{ed1}
F_N(t)=<\chi_{\,\{h_N\,\leq \,t\}\,}>.
\ee
The following expression for the Laplace transform, or characteristic function, of $F_{N}$ is easily obtained from
(\ref{partitionfct}) and (\ref{thermal}):
\be
\label{id}
\varphi_{N} (\lambda)=\int e^{-\lambda t}\, dF_{N}(t) = <e^{-\lambda h_N}> = {Z_N\left(\beta
+{\lambda\over N}\right)\over Z_N(\beta)}\, \cdot
\ee
On the other hand, by Proposition \ref{cwpf} and the mean value theorem we have that, for
$0\leq \beta <1$ and $N$ large enough
\be
\label{id1}
{Z_N\left(\beta
+{\lambda\over N}\right)\over Z_N(\beta)} = \exp{\left( {\lambda\over N}G'_N(\beta^*)\right)}=1+{\cal
O}\left({\lambda\over N}\right)
\ee
for some $\beta^*$ such that $0\leq \beta^*-\beta\leq \lambda/N$. We may now use a well known theorem of
probability theory (see e.g. \cite{S}) which says that $F_N$ converges weakly to $F$ if and only if
$\varphi_N(\lambda) \to \varphi (\lambda)$ for any $\lambda$ (where $\varphi (\lambda)$
is the characteristic function of $F$). Noting that
$\varphi (\lambda) =1$ is the characteristic function of the distribution function
$G(t)={\chi}_{[0,\infty)}(t)$ we have obtained the following result
\vsni
\begin{proposition}
\label{delta} For $0\leq \beta <1$ and $N\to \infty$ the energy densities
$h_N$ converge
in distribution to a random variable $h$ which is
$\delta$-distributed at $x=0$.
\end{proposition}
Moreover, since the range of $h$ is the interval $[-1/2,1/2N]$, the random
variables $h_N^n$ are uniformly integrable for each
$n\in\N$, i.e. for some $\epsilon >0$
\be
\label{uniint}
\sup_N <|h_N|^{n +\epsilon}> \;\; < \; \infty.
\ee
By another well known theorem of probability theory (see \cite{S}) the bound (\ref{uniint})
along with Proposition \ref{delta} imply that
\be
\label{A}
<h_N^n> \; \to \; <h^n> ,\qquad N\to \infty,\quad  n \in \N.
\ee
But we can say more. Indeed, by virtue of (\ref{uniint}) the expansion of
the function $\varphi_{N} (\lambda)$ in powers of $\lambda$,  i.e.
\be
\varphi_{N} (\lambda) =  \sum_{n=0}^\infty (-1)^n {\lambda^n\over n!} <h_N^n>,
\ee
converges absolutely in a domain of the complex $\lambda$-plane which contains the point $\lambda =0$
and can be taken independent of
$N$. Therefore, a standard Cauchy-type estimate along with (\ref{id1}) yield
\be
\label{infinitesimo}
|<h_N^n>- <h^n>| = {\cal O}\left({ C\over N}\right)
\ee
where $C$ is a positive constant depending on $n$ but not on $N$.
On the other hand we have, as $N\to \infty$,
\be
\label{B}-<h_N>= {1\over N^2}\sum_{i<j}<\s_i \s_j> ={N(N-1)\over 2N^2}<\s_1 \s_2>\; \to \; {<\s_1 \s_2>\over
2}.
\ee
Moreover, when computing
$$
<h^2_N>= {1\over N^4}\sum_{i<j,\;\, l<k}<\s_i \s_j \s_l \s_k> ,
$$
the only terms which survive in the limit $N\to \infty$ are those having all the indices distinct.
These can be further divided
into six classes corresponding to the ordered quadruples $i<j<l<k$, $i<l<j<k$, $i<l<k<j$ plus those obtained by
interchanging $i$ with $l$ and $j$ with $k$. Each class contains $N(N-1)(N-2)(N-3)/4!$ terms. Hence we get
\be
\label{C}
<h_N^2>  \; \to \; {<\s_1 \s_2 \s_3 \s_4> \over 4},\quad N\to \infty.
\ee
Now by Proposition \ref{delta} we have that $<h>=0$ and $\V h =<h^2>-<h>^2=0$. Putting together these facts and (\ref{A}),
(\ref{B}), (\ref{C}) we recover the factorization rule (\ref{4pts}).
Notice that we can write
$$
<h_N> =-{1\over N}\left({\partial\log Z_N\over \partial
\beta}\right)=-{1\over N^2}\sum_{i<j}<\s_i \s_j> \; .
$$
Similarly,
\begin{eqnarray}
\V h_N &=& \, <h^2_N> - < h_N>^2
= {1\over
N^2}\left({\partial^2\log Z_N\over \partial \beta^2}\right)\nonumber \\
&=& {1\over N^4}\sum_{i<j,\; l<k}<\s_i \s_j, \s_l \s_k>_c,
\nonumber
\end{eqnarray}
where
\be
<\s_i \s_j ,\s_l \s_k>_c\; :=\; <\s_i \s_j \s_l \s_k>-<\s_i\s_j><\s_l\s_k>.
\ee
More generally, for $n<N/2$ we can write the $n$-th {\sl moment} of $h_N$ as
\begin{eqnarray}
< h^n_N > &=&{(-1)^n\over Z_N\, N^n}\left({\partial^n Z_N\over
\partial
\beta^n}\right) \\
&=& {(-1)^n\over N^{2n}}\sum_{i_1<j_1,\cdots , i_n<j_n}<\s_{i_1} \s_{j_1} \cdots \s_{i_n}
\s_{j_n}>.
\nonumber
\end{eqnarray}
The $n$-th {\sl cumulant} is then defined as
\begin{eqnarray}\label{cumulant}
< h^n_N >_c &:=&{(-1)^n\over N^n}\left({\partial^n\log Z_N\over
\partial
\beta^n}\right) \\
&=& {(-1)^n\over N^{2n}}\sum_{i_1<j_1,\cdots , i_n<j_n}<\s_{i_1}\s_{j_1}, \cdots ,\s_{i_n} \s_{j_n}>_c \; ,
\nonumber
\end{eqnarray}
where $<\s_{i_1}\s_{j_1}, \cdots ,\s_{i_n} \s_{j_n}>_c$ is called the {\sl pairwise $n$-points
connected} (or {\sl truncated}) correlation, and is defined recursively by
\begin{eqnarray}
\label{connected}
<\s_{i_1}\s_{j_1}, \cdots ,\s_{i_n} \s_{j_n}>_c &=& <\s_{i_1}\s_{j_1}
\cdots \s_{i_n} \s_{j_n}>
\\
\nonumber
&-& \sum_{  {\rm partitions} \; {\rm of}\atop i_1 j_1,\dots ,i_n
j_n}\left[\matrix{\hbox{products of pairwise $m$-points}\hfill \cr
\hbox{connected correlations with} \; m<n
\hfill\cr } \right]\cdot
\end{eqnarray}
While the moments $< h^n_N >$ are somehow redundant in that they carry information on correlations among $k$ spins
with $k\leq n$ (so that part of this information is already stored in lower order moments),
the cumulants $< h^n_N >_c$
carry only the new information concerning $n$ spins.

Using once more the fact that $h^n_N$ is uniformly integrable, Proposition
\ref{delta}
and (\ref{infinitesimo}) we see that
$< h^n_N >_c = {\cal O}( C\, N^{-1})$ for each fixed $n\in \N$ and $N\to \infty$.

We have therefore proved the following
\begin{proposition}
\label{conn} For $0\leq \beta <1$ and for each $n>1$ we can find a
positive
constant $C=C(n)$ so that
\be
\label{conn2}
<\s_{i_1}\s_{j_1}, \cdots ,\s_{i_n} \s_{j_n}>_c\, =  \,{\cal O}\left({ C\over N}\right)
\ee
as $N\to \infty$.
\end{proposition}
\begin{remark}
By a straightforward inductive argument based on the recursive
formula (\ref{connected}) and on translation invariance it is easily seen that in the
thermodynamic limit (\ref{conn2}) amounts to the simple factorization property
\be
\label{conn3}
<\s_1\s_2 \cdots \s_{2n-1} \s_{2n}>\, = \, <\s_{1}\s_{2} >^n \, .
\ee
\end{remark}


\section{The orthogonal model}
We shall now extend the results obtained in the previous Section to the more
interesting class of (non-translationally invariant) interactions

\be
\label{hamiltonian}
{\cal H}_N(\s)=-\frac{1}{2}\sum_{i,\, j \in {\bf Z}_N^2}J_{ij}\,\s_i\, \s_j
\ee
where ${\bf Z}_N$ is the integer lattice ${\bf Z}_N={\bf Z}\,{\rm mod}\, N$ and $J=(J_{ij})$ is a
symmetric real orthogonal $N\times N$ matrix. This means that $J$ has
 the form
$J=OLO^T$ with $L$ a diagonal matrix with elements
$\pm 1$ and $O$ a generic orthogonal matrix chosen at random w.r.t. the Haar measure on the orthogonal
group. The knowledge of the eigenvalues of $J$ imposes simple bounds on the energy of
any spin configuration.
Here, due to orthogonality, the possible eigenvalues are $+1,-1$
so that
\be
 -\frac{N}{2} \leq {\cal H}_N(\s) \leq  \frac{N}{2} \cdot
\ee
An important example
for our purposes is given by the {\it sine model} where
\be
J_{i,j}=\frac{2}{\sqrt{2N+1}}\sin{\left(\frac{2\pi ij}{2N+1}\right)}
\label{}
\ee
which satisfies (see e.g.\cite{MPR,DEGGI}):
\begin{eqnarray}
\label{orth}
J \, J^{T} &=& {\rm Id}
\\
\label{traccia1} \sum_{i=1}^N J_{ii}&=&1\quad\hbox{for}\quad N \quad
\hbox{odd};
\\
\label{traccia0}  \sum_{i=1}^N J_{ii}&=&0\quad \hbox{for}\quad N \quad
\hbox{even}; \\
\sum_{i=1}^N
J_{ii}^2&=&1.
\end{eqnarray}
 the last equation being  a particular  Gauss sum (see e.g.
\cite{Ap}).
\begin{remark}
One might also consider interaction matrices with zero diagonal terms,
recovering orthogonality in large $N$ limit. This amounts  to consider the shifted Hamiltonian
\be
{\tilde {\cal H}}_N(\s)={\cal H}_N(\s)+{1\over 2},
\ee
so that the average energy is equal to zero
(instead of $-1/2$),
and may be convenient for particular purposes (see also Remark \ref{cw}).
\end{remark}

\subsection{Mean field properties at any temperature: the second order}
The results of this section hold for the particular choice of the  {\it sine
model}
\be
J_{i,j}=\frac{2}{\sqrt{2N+1}}\sin{\left(\frac{2\pi ij}{2N+1}\right)} \; .
\label{}
\ee
defined above.
\begin{proposition}
\label{squarefluctuation}
Denote $D_r(N)$ the {\rm fat diagonal} of dimension $r$, i.e.
the set of points of ${\bf Z}_N^r$
in which at least two of the indices coincide. Let the bar indicate the
complementary set $\overline{D_r(N)}$. For every positive inverse temperature
$\beta$ except, at most, a set of zero Lebesgue measure, the Gibbs state
$<\;>$ expectations fulfill the following relation:
\begin{eqnarray}
\label{wicklikeformula}
\biggl|\, \frac{1}{N^2}\sum_{i,\, j,\, l,\, m\, \in \overline{D_4(N)}} &&J_{i,j}J_{l,m}
<\sigma_i\sigma_j\sigma_l\sigma_m>
- \nonumber \\
&&- \;\frac{1}{N^2}\sum_{i,\, j\in \overline{D_2(N)};\; l,\, m\in \overline{D_2(N)}
}J_{i,j}J_{l,m}
<\sigma_i\sigma_j><\sigma_l\sigma_m>\, \biggr|\; = \; {\cal O}\left(\frac{1}{N}\right)\nonumber
\end{eqnarray}
\end{proposition}
\begin{remark}
The previous formula is {\it homogeneous} in the following sense:  thanks to
restriction outside the fat diagonal, the first term only contains $4$-points
correlations, and the second terms only products of  $2$-points correlations.
In this  sense it can be considered as the natural generalization of the
simple factorization property valid for  the Curie-Weiss case.
\end{remark}
 Conceptually the
proof relies on the equivalence of the microcanonical and canonical ensemble
which in one of its formulations says that the energy density has vanishing
fluctuations with respect to the Gibbs measure in the thermodynamical limit.

The theorem is structured in several lemmata:
\begin{lemma}
\label{equivensemble}
For every positive temperature $\beta$ but at most a set of zero
Lebesgue measure, the internal energy density has zero quadratic
fluctuations, i.e.
\be
<h_N^2>-<h_N>^2={\cal O}\left(\frac{1}{N}\right) \; .
\ee
\label{}
\end{lemma}
{\sl Proof.} By the definition of free energy density
\be
-\beta f_N(\beta) \; = \; \frac{1}{N}\log \sum_{\sigma}\exp^{-\beta {\cal H}_N (\sigma)}
\; ,
\ee
\be
\frac{d}{d\beta}(\beta f_N(\beta))\; = \; <h_N> \; ;
\ee
\be
\label{seconderi}
\frac{d^2}{d\beta^2}(\beta f_N(\beta))\; = \; - N (<h_N^2>-<h_N>^2) \; .
\ee
The function $-\beta f_N(\beta)$ is bounded and convex
with  bounded derivative
(the boundedness comes from the bounds on the Hamiltonian \cite{DEGGI} due to
orthogonality and convexity, and is proved on very general  grounds in
\cite{R}). The  $N\to \infty$ limit of $-\beta f_N(\beta)$, which again by
convexity always exists, at least along subsequences \cite{R}, is itself
convex and has always right and left derivatives which coincide except
at most on a countable set of points. Integrating the (\ref{seconderi})
in any $\beta$ interval the positivity of the left hand side and
the fundamental theorem of calculus yield the lemma.

\begin{lemma}
\label{orthogonal1}
Case $i=j$; for the sine interaction $J$ the following result holds:
\be
\left|\frac{1}{N^2}\sum_{i,l,m}J_{i,i}J_{l,m}
<\sigma_l\sigma_m>\right| \; \le \; \frac{1}{2N}
\ee
\label{}
\end{lemma}
{\sl Proof.} We have
\be
\left|\frac{1}{N^2}\sum_{i,l,m}J_{i,i}J_{l,m}
<\sigma_l\sigma_m>\right|\leq \left|\frac{1}{N^2}\sum_{l,m}J_{l,m}
<\sigma_l\sigma_m>\right|\le  \frac{1}{N^2}\cdot \frac{N}{2}={1\over 2N} \, ,
\label{}
\ee
where the first inequality comes from
$\sum_{i}J_{i,i}\leq 1$ which in turn is a consequence of (\ref{traccia1})-(\ref{traccia0}).  The
second inequality is true because the maximum of the Hamiltonian $\displaystyle\sum_{l,m}J_{l,m}
\sigma_l\sigma_m$, which is an upper bound for its expectation, is $N/2$.
\begin{lemma}
\label{orthogonal2}
Case $j=l$; for the sine interaction $J$ the following result holds:
\be
\frac{1}{N^2}\sum_{i,j,m}J_{i,j}J_{j,m}
<\sigma_i\sigma_m> \; = \; \frac{1}{N}
\ee
\label{}
\end{lemma}
{\sl Proof.}
$$
\frac{1}{N^2}\sum_{i,j,m}J_{i,j}J_{j,m}
<\sigma_i\sigma_m> \; = \; \frac{1}{N^2}\sum_{i,m}\delta_{i,m}
<\sigma_i\sigma_m> \; = \; \frac{1}{N^2}\sum_{i} 1 \; = \;  \frac{1}{N} \; \;\;.
\qed
$$
\noindent
{\sl Proof of Proposition \ref{squarefluctuation} }\\
\noindent
Defining
$$
A \; = \; \left|\frac{1}{N^2}\sum_{i,l,m}J_{i,i}J_{l,m}
<\sigma_l\sigma_m>\right| \; ,
$$
and
$$
B \; = \; \left| \frac{1}{N^2}\sum_{i,j,m}J_{i,j}J_{j,m}
<\sigma_i\sigma_m> \right|\; ,
$$
we have
\begin{eqnarray}
\biggl|\frac{1}{N^2}\sum_{i,\, j,\, l,\, m\, \in \overline{D_4(N)}}J_{i,j}J_{l,m}
&<&\sigma_i\sigma_j\sigma_l\sigma_m>
-\nonumber \\ \ &&\frac{1}{N^2}\sum_{i,\, j\in \overline{D_2(N)};\;
l,\, m\in \overline{D_2(N)} }J_{i,j}J_{l,m}
<\sigma_i\sigma_j><\sigma_l\sigma_m>\biggr| \nonumber \\
&&\leq \; \;<h_N^2>-<h_N>^2 + 6A +4B \;. \nonumber
\end{eqnarray}
The previous lemmata provide the claim. $\qed$

\subsection{High temperature expansion of the free energy for the
orthogonal model}
We can now try to mimic the procedure used in the previous section to
decouple the spins. To this end, let $B$ be an
orthogonal matrix such that
$B^TJB=D$ with
$D={\rm diag}\, (d_1,\dots ,d_N)$.
Since ${\rm det}\, J\ne 0$ we have $d_i>0$, $i=1,\dots ,N$, and
${\rm det}\, J^{-1}=\prod_{i}d_i^{-1}$.
Let $u\in \R^N$ be such that $\s = B u$.
We have
$<J\s,\s>=<Bu,JBu>=<u,Du>$, and thus
\begin{eqnarray}
\exp({\lambda\over 2N}<J\s,\s>)&=&\prod_{i=1}^N\exp({\lambda\over 2N}d_iu_i^2 )
\nonumber \\
&=&\prod_{i=1}^N{1\over
\sqrt{2\pi}}
\int_{-\infty}^\infty \exp\left(-{x_i^2\over 2}+
\sqrt{{\lambda d_i\over N}}\,\, u_ix_i\right) dx_i \nonumber \\
&=&{1\over (2\pi)^{N/2}}\int_{\R^N}
\exp\left(-{1\over 2}<x,x>+
\sqrt{{\lambda \over N}}\,\, <u, D^{1/2}x>\right) \, dx\nonumber \\
&=&{{\rm det}\, J^{-{1\over 2}}\over (2\pi \lambda)^{N/2}}\int_{\R^N}
\exp\left(-{1\over 2\lambda}<y,J^{-1}y>+
\, <\s, {y\over \sqrt{N}}>\right) \, dy.\nonumber
\end{eqnarray}
By (\ref{partitionfct}) this yields
\be
Z_N(\beta) = 2^N
{{\rm det}\, J^{-{1\over 2}}\over (2\pi \beta)^{N/2}}\int_{\R^N}
\exp \biggl(-{1\over 2\beta}<y,J^{-1}y>+
+\sum_i \log \cosh {y_i\over \sqrt{N}}\biggr) \, dy \nonumber
\ee
to be compared with (\ref{logcosh}). We point out that the square roots
appearing in the above formula are only apparently
ill defined. Indeed they disappear as soon as one takes its development in powers of $\beta$, because the latter
contains only even terms. $Z_N(\beta)$  has been computed by Parisi and Potters
in \cite{PP} using standard high-temperature techniques. Relying on their
computation we are now in the position to
state a result analogous to Proposition
\ref{cwpf} for this class of models.
\begin{proposition}
\label{ompf}
For $0\leq \beta <1$ and for any orthogonal interaction we have
\be\label{freen}
-\beta\; {F(\beta)}\equiv\log Z_N (\beta) =N\log 2\, + N G_N(\beta),
\ee
where
\be
\label{convergence}
G_N(\beta) \nearrow
G(\beta) = {1\over 4}\left[ \sqrt{1+4\beta^2}-\log\left({1+\sqrt{1+4\beta^2}\over 2}\right)-1\right]
\quad\hbox{as}\quad N\to \infty.
\ee
\end{proposition}

\subsection{Limiting behaviour and connected
correlations at high temperature for the orthogonal model}

We start noticing that the function $G(\beta)$ defined in Proposition \ref{ompf} has the following expansion in the
vicinity of $\beta=0$:
\be
G(\beta) = {\beta^2\over 4} + {\cal O}(\beta^3)
\ee
which, by the way, coincides with what one obtains for the SK model if
truncated after the first term. Moreover,
according to Proposition \ref{ompf} and with the same notation of the previous section, we have
\be
\label{id2}
<e^{-\lambda h_N}>={Z_N\left(\beta
+{\lambda\over N}\right)\over Z_N(\beta)} = \exp{\left( {\lambda}\, G'_N(\beta^*)\right)}
\ee
for some $\beta^*$ such that $0\leq \beta^*-\beta\leq \lambda/N$.
We now observe that according to (\ref{freen}) we have
\be
<h_N> = -{1\over N} \left({\partial \log Z_N\over \partial \beta}\right) = - G_N'(\beta)
\ee
and since $<h_N>$ is bounded uniformly in $N$ property (\ref{convergence}) implies
\be
<h_N> \; \to \; <h> \, = \, - G'(\beta) = - {\beta \over 1+ \sqrt{1+4\beta^2}}\quad\hbox{as}\quad N\to \infty.
\ee
Therefore, if we fix $\beta \in [0,1)$ and expand the r.h.s.
of (\ref{id2}) in a neighborhood of $\beta^{\ast}$ we obtain for
$N$ large enough
\be
<e^{-\lambda h_N}> = e^{\lambda\, G'(\beta)}\left(1+{\cal O}
\left(\lambda^2\over N\right)\right).
\ee
Note that $G'(0)=0$, so that at infinite temperature ($\beta=0$) we recover the same result as in Proposition
\ref{delta} for this class of models (see also
\cite{DEGGI}, Section 3).
More generally we have the following,
\begin{proposition}
\label{delta1} For $0\leq \beta <1$ and $N\to \infty$ the energy densities
$h_N$ converge
in distribution to a random variable $h$ which is
$\delta$-distributed at $x=-G'(\beta)$.
\end{proposition}
Mimicking again the argument of the Curie-Weiss case we introduce
the $n$-th moment
\begin{eqnarray}
< h^n_N > &=&{(-1)^n\over Z_N\,N^n}\left({\partial^n Z_N\over
\partial
\beta^n}\right) \\
&=& {(-1)^n\over (2N)^{n}}\sum_{i_1,\cdots , i_n \in {\bf Z}_N^n \atop j_1,\cdots , j_n \in {\bf Z}_N^n}
J_{i_1j_1}\cdots J_{i_nj_n}<\s_{i_1} \s_{j_1}\cdots \s_{i_n} \s_{j_n}>
\nonumber
\end{eqnarray}
and the $n$-th cumulant
\begin{eqnarray}
< h^n_N >_c &=&{(-1)^n\over\,N^n}\left({\partial^n \log Z_N\over
\partial
\beta^n}\right) \\
&=& {(-1)^n\over (2N)^{n}}\sum_{i_1,\cdots , i_n \in {\bf Z}_N^n \atop j_1,\cdots , j_n \in {\bf Z}_N^n}
J_{i_1j_1}\cdots J_{i_nj_n}<\s_{i_1} \s_{j_1}, \cdots ,\s_{i_n} \s_{j_n}>_c
\nonumber
\end{eqnarray}
where $<\s_{i_1} \s_{j_1}, \cdots ,\s_{i_n} \s_{j_n}>_c$ is defined as above
(see (\ref{cumulant}) and (\ref{connected})). We may now use Proposition \ref{delta1} to conclude
that the cumulants vanish in the thermodynamic limit. However, at variance
with the (translationally invariant) Curie-Weiss model where summing over the
indices $i_1,\cdots , i_n, j_1,\cdots , j_n$ produces only an overall
combinatorial factor, so that the vanishing of the $n$-th cumulant can be
immediately translated into the vanishing of  pairwise $n$-points connected
correlations, here we have to be content with the following result.
\begin{proposition}
\label{factor} For any orthogonal interaction and
$0\leq \beta <1$ we can find
a positive constant $C=C(n)$ so that
as $N\to \infty$
\be
{1\over N^{n}}\sum_{i_1,\cdots , i_n \in {\bf Z}_N^n \atop j_1,\cdots , j_n \in {\bf Z}_N^n}
J_{i_1j_1}\cdots J_{i_nj_n}<\s_{i_1} \s_{j_1}, \cdots ,\s_{i_n} \s_{j_n}>_c \; = {\cal O}
\left({C\over
N}\right).
\label{facfor}
\ee
\end{proposition}
We are now in position to strengthen, for the sine interaction, the previous
proposition into a factorization-like formula by showing that for each
expectation only the terms outside the {\it fat diagonal} contribute
to the sum. For this purpose we can prove the following:
\begin{proposition}
\label{factorw}
For the sine interaction $0\leq \beta <1$ we can find
a positive constant $C=C(n)$ so that
as $N\to \infty$
$$
{1\over N^{n}}
\sum{}^{\ast}\,
J_{i_1j_1}\cdots J_{i_nj_n}<\s_{i_1} \s_{j_1}, \cdots ,\s_{i_n} \s_{j_n}>_c \; = {\cal O}
\left({C\over
N}\right)
$$
where the starred sum means the following: first apply to the pairwise
$n-$points connected correlations the definition (\ref{connected}); then for
each of the resulting terms sum only over  distinct indices within
each expectation.
\end{proposition}
{\it Example}:  Let $n=3$. Then:
\begin{eqnarray*}
&&\sum{}^{\ast}\,
J_{i_1j_1}J_{i_2j_2}J_{i_3j_3}<\s_{i_1}\s_{j_1},\s_{i_2}\s_{j_2},
\s_{i_3} \s_{j_3}>_c =
\\
&=& \sum_{i_1,i_2, i_3, j_1,j_2, j_3\in \overline{D_6(N)}}
J_{i_1j_1}J_{i_2j_2}J_{i_3j_3}<\s_{i_1}\s_{j_1}\s_{i_2}\s_{j_2}
\s_{i_3} \s_{j_3}>
\\
&-&\sum_{i_1,i_2,j_1,j_2\in \overline{D_4(N)}\atop i_3,j_3\in
\overline{D_2(N)}}
J_{i_1j_1}J_{i_2j_2}J_{i_3j_3}<\s_{i_1}\s_{j_1}\s_{i_2}\s_{j_2}>< \s_{i_3}
\s_{j_3}>
\\
&-&\sum_{i_1,i_3,j_1,j_3\in \overline{D_4(N)}\atop i_2,j_2\in
\overline{D_2(N)}}
J_{i_1j_1}J_{i_2j_2}J_{i_3j_3}<\s_{i_1}\s_{j_1}\s_{i_3}\s_{j_3}>< \s_{i_2}
\s_{j_2}>
\\
&-&\sum_{i_2,i_3,j_2,j_3\in \overline{D_4(N)}\atop i_1,j_1\in
\overline{D_2(N)}}
J_{i_1j_1}J_{i_2j_2}J_{i_3j_3}<\s_{i_2}\s_{j_2}\s_{i_3}\s_{j_3}>< \s_{i_1}
\s_{j_1}>
\\
&+& 2\sum_{i_1,j_1\in
\overline{D_2(N)}\atop i_2,j_2\in\overline{D_2(N)}; i_3,j_3\in
\overline{D_2(N)}}
J_{i_1j_1}J_{i_2j_2}J_{i_3j_3}<\s_{i_2}\s_{j_2}><\s_{i_3}\s_{j_3}>< \s_{i_1}
\s_{j_1}>
\end{eqnarray*}
{\sl Proof of Proposition \ref{factorw}.} The proof is by induction over
$n\ge 2$.
Set first $n=2$. Then the proof follows from \ref{factor} and  lemmata
\ref{orthogonal1} and
\ref{orthogonal2} by the same argument of Proposition
\ref{squarefluctuation}.
The inductive argument then proceeds as follows:
for each $n$ Proposition 3.9
yields  a quantity of order $1/N$.
To prove Proposition 3.10 we have to show that the contribution
to (\ref{facfor}) coming from each term belonging to the fat diagonal
vanishises at least as $1/N$ as $N\to\infty$. We prove this last
part by induction. Since in the fat diagonal at least two indices
coincide we separate two cases:\\

1) The two indices belong to the same pair; for instance,
$i_1=j_1$;

2) The two indices belong to different pairs.

Let us first consider case 1. Here the l.h.s. of
(\ref{facfor}) becomes
$$
P_1(N):={1\over N^{n}}\sum_{i_1,\cdots , i_n \in {\bf Z}_N^n \atop
i_1,\cdots , j_n \in {\bf Z}_N^n}
J_{i_1i_1}\cdots J_{i_nj_n}<\s_{i_1} \s_{i_1}, \cdots ,\s_{i_n}
\s_{j_n}>_c
$$
However $\s_{i_1} \s_{i_1}=1$; hence the sum over $i_1$ can be performed.
Since $\ds \sum_{i}J_{ii}$ is either $0$ or $1$  the result is either
$P_1(N)=0$ or
$$
P_1(N)={1\over N^{n}}\sum_{i_2,\cdots , i_n \in {\bf Z}_N^n \atop
j_1,\cdots , j_n \in {\bf Z}_N^n}
J_{i_2i_2}\cdots J_{i_nj_n}<\s_{i_2} \s_{i_2}, \cdots ,\s_{i_n}
\s_{j_n}>_c
$$
Therefore, in any case we get the estimate
\begin{eqnarray*}
|P_1(N)|\leq \frac{1}{N}\frac{1}{N^{n-1}}\sum_{i_2,\cdots , i_n \in {\bf Z}_N^n \atop
j_1,\cdots , j_n \in {\bf Z}_N^n}
J_{i_2i_2}\cdots J_{i_nj_n}<\s_{i_2} \s_{i_2}, \cdots ,\s_{i_n}
\s_{j_n}>_c ={\cal O}\left(\frac{1}{N}\right)
\end{eqnarray*}
because we can apply (\ref{facfor}) for the $(n-1)$-points connected
correlations. The same arguments applies to any other term $P_k(N):
k=2,\ldots,n$ where obviously $P_k(N)$ is the l.h.s. of (\ref{facfor})
with $i_k=j_k$.

There remains case 2. Here there will be either terms
in $<\s_{i_1} \s_{j_1}, \cdots ,\s_{i_n} \s_{j_n}>_c$ for which the two equal indices
(for example $i_1$ $i_2$) appear within the same expectations or terms where
they appear in different ones. For the former we observe that the identity $\s_{i_1}\s_{i_2}=1$
reduces us to a pairwise $(n-1)$-points connected correlation. For the latter the
use of (\ref{connected}) where the right hand side summation is extended
only to those partitions that keep the indices $i_1$ and $i_2$ in different expectations allows us
to use the inductive hypothesis (since they are just products of pairwise
connected correlations of lower order). Therefore we can factor a term of order $1/N$ out of
each expectation, thus concluding the proof. $\qed$ \\
{\it Example: case n=3}. Then by the formula (\ref{connected}) and $\s^2=1$:
\begin{eqnarray*}
&& <\s \s_{j_1},\s \s_{j_2}, \s_{i_3} \s_{j_3}>_c =
\\
&&
<\s_{j_1}\s_{j_2},\s_{i_3}\s_{j_3}>_c - <\s\s_{j_1},\s_{i_3}\s_{j_3}>_c<\s\s_{j_2}>_c -
<\s\s_{j_2},\s_{i_3}\s_{j_3}>_c<\s\s_{j_1}>_c \; .
\end{eqnarray*}

\vskip 0.3cm\noindent

{\bf Acknowledgments} One of us (PC) 
thanks Francesco Guerra for many enlightening
discussions. All authors would like to thank two referees for pointing
out a mistake and many useful suggestions for improving the presentation.

\end{document}